\documentclass[final,5p,times]{elsarticle}

\usepackage{graphicx}

\usepackage{amssymb}

\usepackage{lineno}
\usepackage{subfigure} 

\biboptions{comma,square}

\journal{Nuclear Instruments and Methods in Physics Research Section A}

\begin{document}

\begin{frontmatter}



\title{The LHCb VELO Upgrade}

\author[pabloRodriguez]{\textsl{Pablo Rodr\'iguez P\'erez} on behalf of \textbf{the LHCb VELO group}}
\address[pabloRodriguez]{University of Santiago de Compostela}

\author{}

\address{}

\begin{abstract}
LHCb is a forward spectrometer experiment dedicated to the study of new physics in the decays of beauty and charm hadrons produced in proton collisions at the Large Hadron Collider (LHC) at CERN.

The VErtex LOcator (VELO) is the microstrip silicon detector surrounding the interaction point, providing tracking and vertexing measurements.
The upgrade of the LHCb experiment, planned for 2018, will increase the luminosity up to $\rm 2\times10^{33}\ cm^{-2}s^{-1}$ and will perform  the readout as a trigger-less system with an event rate of 40 MHz. 
Extremely non-uniform radiation doses will reach up to $\rm 5~\times~10^{15}$~1~MeV$\rm n_{eq}/cm^2$ in the innermost regions of the VELO sensors, and the output data bandwidth will be increased by a factor of 40.
An upgraded detector is under development based in a pixel sensor of the Timepix/Medipix family, with 55~x~55~$\rm \mu m^2$ pixels. 
In addition a microstrip solution with finer pitch, higher granularity and thinner than the current detector is being developed in parallel.

The current status of the VELO upgrade program will be described together with recent testbeam results.
\end{abstract}


\end{frontmatter}

\linenumbers


  \section{The VELO detector at the LHCb experiment}

    \subsection{The LHCb experiment}
      The LHCb \cite{lhcb} experiment is designed to exploit the enormous production cross sections of b and c hadrons in proton collisions at the LHC.
      It has demonstrated an excellent momentum resolution and particle identification capabilities, needed to study rare B decays.
      As the $\rm b\overline{b}$ pairs are produced predominantly in the forward and backward directions the experiment was built to cover the pseudo-rapidity range of $\rm1.9~<~\eta~<~4.9$.
      LHCb has been running from September 2009 and it has recorded 1.1~$\rm~fb^{-1}$ pp collisions during 2011 and 2.1~$\rm~fb^{-1}$ so far during 2012.
      Currently LHCb is running at a luminosity of \newline $\rm \mathcal{L}~=~4~\times~10^{32}~cm^{-2}~s^{-1}$ with an average pile up of 1.4 interactions per bunch crossing (Fig.~\ref{subfig:avLumi}).
      Both values are above design specifications which is an indication of the excellent performance of the experiment.
  
    \subsection{The VErtex LOcator (VELO)}
      \begin{figure}[tb]
       
      \subfigure[Average luminosity]{
	  \includegraphics[bb=0 0 200 125]{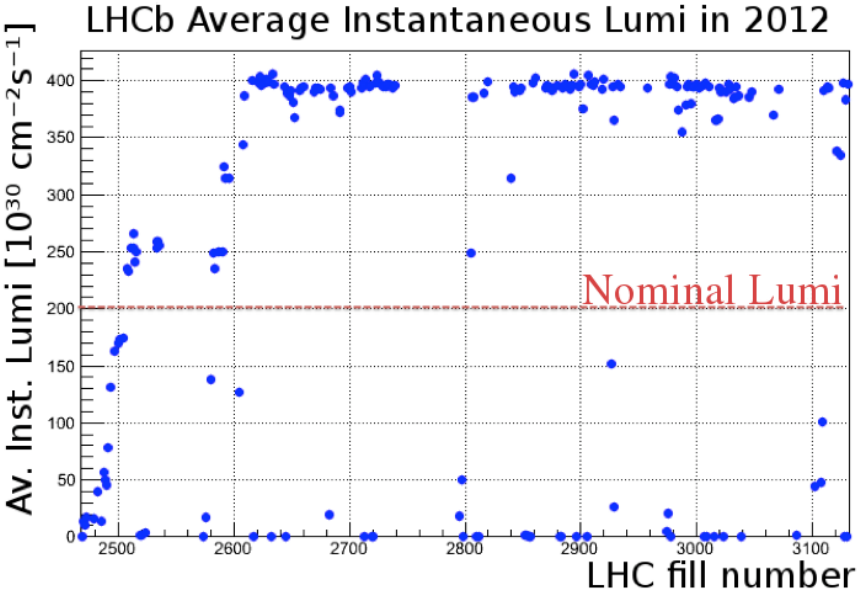}
	\label{subfig:avLumi}
      }
      \subfigure[CAD image of the VELO detector]{
	\includegraphics[bb=0 0 200 200]{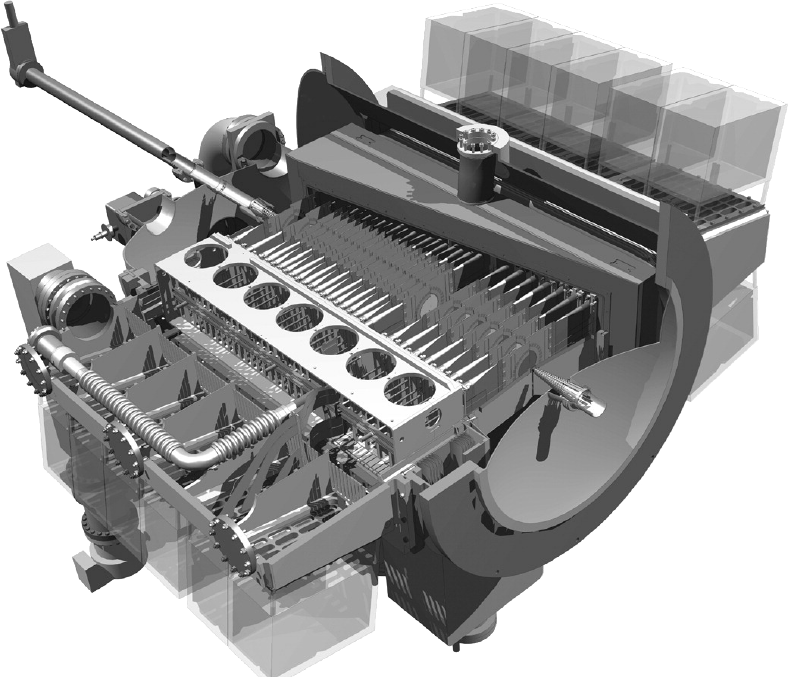}
	\label{subfig:velo_modules}
      }
      \caption{Fig.~\ref{subfig:avLumi} compares average instantaneous luminosity in the LHCb experiment with the nominal value. Fig.~\ref{subfig:velo_modules} shows 3D representation of the current VELO detector}
      \end{figure}
      
      The VErtex LOcator \cite{vanBeuzekom200821} is the silicon detector surrounding the LHCb interaction point.
      It is located only 7~mm from the beam axis during normal operation and measures very precisely the primary vertex and the decay vertices.
      It consists of two mobile detector halves with 21 modules each (Fig.~\ref{subfig:velo_modules}). 
      Every module is composed of two $\rm n^+$-on-n 300~$\rm \mu$m thick microstrip sensors, with a pitch range from 38~$\rm \mu$m to 102~$\rm \mu$m providing R and $\rm \phi$ measurements.
      The 2048 channels of each sensor are readout at 1~MHz with 16 analogue front-end chips named Beetles~\cite{beetle}.  
      Each sensor has its own hybrid circuit, and it is glued onto a thermal pyrolytic graphite (TPG) foil which provides a thermal highway for the module.
      The TPG foil is coated with carbon fibre to give mechanical support to the module.
      The cooling system is based on a bi-phase $\rm CO_2$ plant which delivers the coolant via stainless steel tubes to the cooling blocks clamped to the base of the hybrids. 
      
      The detector is placed inside the beam pipe and each half is separated from the LHC beams by a vacuum box named RF-foil.
      The RF-foil is the corrugated AlMg3 foil of 300~$\rm \mu$m that is between the VELO sensors and beams. 
      It encapsulates the VELO modules in a secondary vacuum and shields the front end electronics from the beams currents.
      
      As the sensors are placed so close, and perpendicular to the beams, the radiation dose is highly non-uniform reaching $\rm 2.5~\times10^{14}~1~MeVn_{eq}/cm^2$ in the innermost regions.
      After almost 3 years of operation, the VELO has already experienced radiation damage~\cite{velo_rad_damage}.
      Leakage current has increased around 1.9~$\rm \mu$A per 100~$\rm pb^{-1}$ delivered, following the delivered luminosity as expected.
      Type inversion has been measured in the inner regions of the sensors, being the first sensors in LHC to show it.
      More detailed information about radiation damage in VELO can be found in \cite{gCasse}.
      Despite the radiation damage no significant performance degradation has been shown in the installed VELO, providing a decay time resolution of $\rm \sim50$~fs and a point resolution of only $\rm 4~\mu$m in the region of $\rm 40~\mu$m pitch. 
      More detailed information about the VELO performance can be found in \cite{kSenderowska}.

  \section{The VELO upgrade}
    \begin{figure}[tb]
    \centering 
    \subfigure[Radiation dose as function of the distance to the beams]{
      \includegraphics[bb=0 0 250 100]{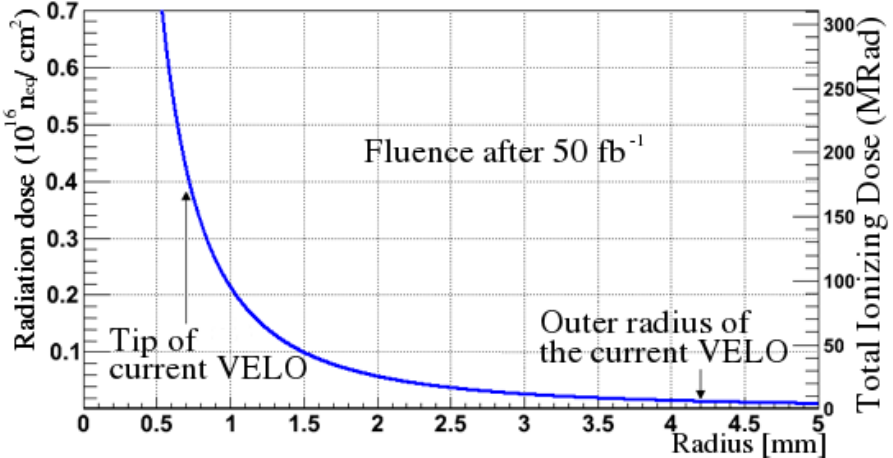}
      \label{subfig:doseVsRadius}
    }
    \subfigure[Rate of tracks per bunch crossing in a pixel based module]{
      \includegraphics[bb=0 0 200 210]{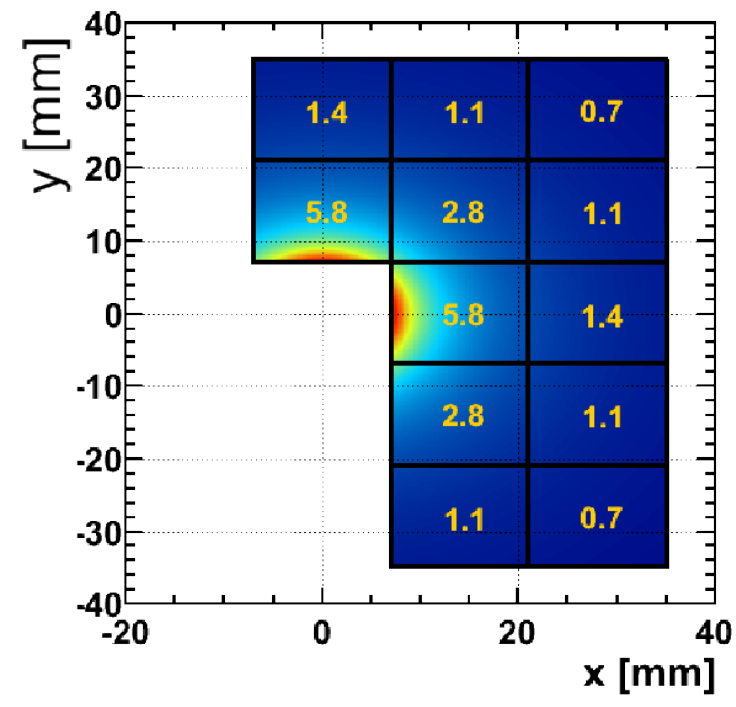}
      \label{subfig:pixel_rate}
    }
    \caption{\small Expected values of radiation and track rate per bunch crossing for the upgraded VELO.}
    \end{figure}  
    
    As it was explained, LHCb is running above design parameters, recording around 2~$\rm fb^{-1}$ per year, expecting up to 9~$\rm fb^{-1}$ by the end of 2017.
    In that moment LHCb will be able to cover its primary physics goals, namely the search for New Physics via the measurement of CP asymmetries and rare decays of b and c quarks.
    Nevertheless, many physics channels will still be statistically limited.
    An upgrade is planned for 2018 and the luminosity will be increased up to $\rm 2\times10^{33}~cm^{-2}s^{-1}$, reaching 10~$\rm fb^{-1}$ per year.
    The radiation dose will reach up to 230 MRad or $\rm 5\times10^{15}~1~MeV~n_{eq}/cm^2$ in the inner regions of the sensors.
    The radiation dose will be extremely non-uniform as can be appreciated in Fig.~\ref{subfig:doseVsRadius}.

    Luminosity by itself will not improve hadronic event yield since the current bottleneck is the hardware trigger.
    Currently the LHCb trigger is implemented in two stages.
    The first level is a hardware implemented trigger, carried out with information from the Muon and Calorimeters detectors.
    The next step is the High Level Trigger (HLT) where more refined selections are made in a CPU-farm using data from the others detectors, like the displaced vertices information provided by the VELO.
    The hardware trigger was implemented to filter the event rate from 40~MHz to a maximum of 1~MHz so the HLT can process them.
    Thus the effective readout rate for the current VELO is $\rm \leq$~1~MHz.
    
    After the upgrade, all the detectors will be readout at 40~MHz, providing data to a fully flexible software trigger, which will select events of interest according to information like the reconstructed vertices.
    This trigger-less system will raise the bandwidth needs by a factor of 40, promoting the LHCb to the category of a general purpose experiment in the forward direction.
    This implies that all the front-end electronics must be redesigned or adapted to cope with the new data rate requirements, implementing new functionalities like zero suppression in order to reduce the bandwidth.
    Preliminary simulations shows an average rate of 26 tracks per sensor per bunch crossing, giving an output rate above 2 Tbit/s from the whole VELO, Fig.~\ref{subfig:pixel_rate}.
    
    \subsection{Sensor technology}
    \begin{figure}[tb]
    \centering 
    \subfigure[Proposed pixel module]{
      \includegraphics[bb=0 0 250 125]{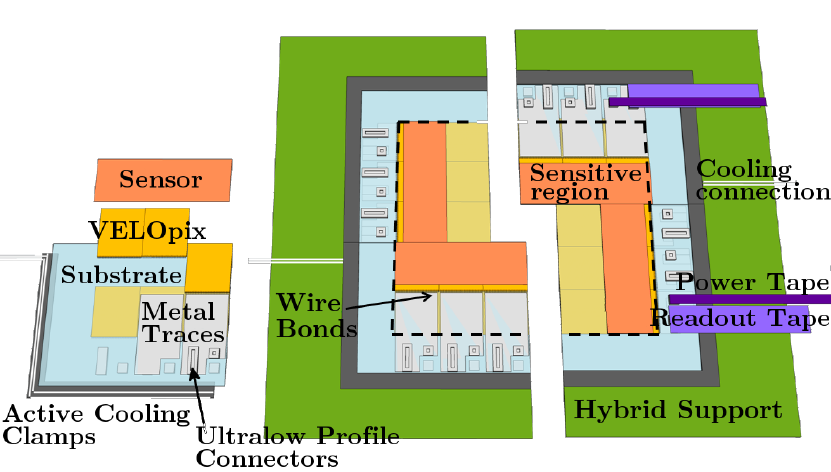}
      \label{subfig:pixel_module}
    }
    \subfigure[Microstrip prototypes with R and $\rm \phi$ geometry]{
      \includegraphics[bb=0 0 250 230]{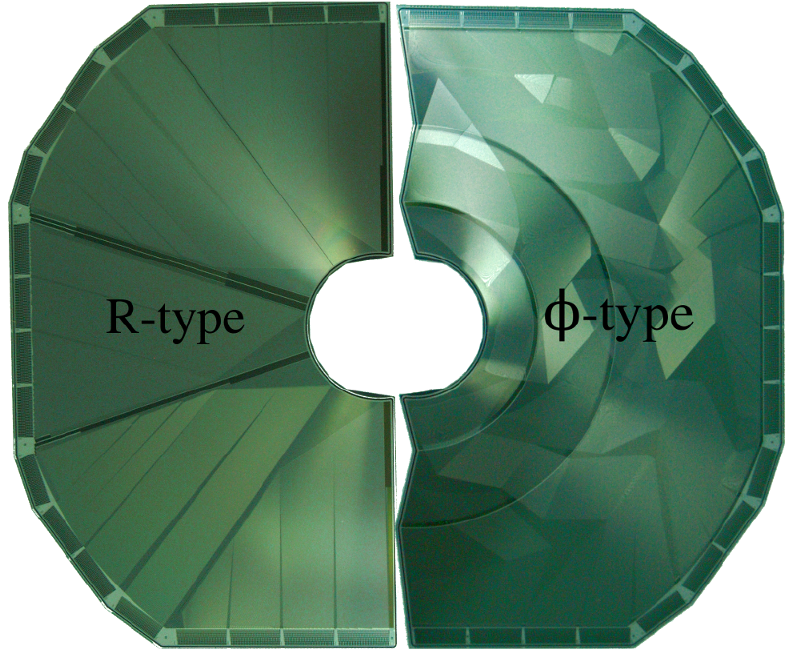}
      \label{subfig:strip_prototype}
    }
    \caption{\small Upgraded VELO modules will be based on pixel of microstrip sensors.}
    \end{figure}  
    
    Two silicon technologies are being considered to meet the requirements of the VELO upgrade: pixels and microstrips. 

    \paragraph{Pixel modules}
      The pixel option is based on an ASIC named VeloPix, from the Medipix/Timepix~\cite{Llopart2007485} family, bump bonded onto a n-on-p sensor.
      The VeloPix has a matrix of $\rm 256\times256$ pixels of $\rm 55\times55~\mu m^2$ each.
      It will provide simultaneously information about the deposited energy and the time stamp, reaching an output rate at the hottest chip $>$~12~Gbit/s.
      More detailed VeloPix specifications are given in Table~\ref{tab:velopix}, and a sketch of a proposed pixel module is shown in Fig.~\ref{subfig:pixel_module}.
      
      \begin{table}
      \begin{center}
      \begin{tabular}{|p{.2\textwidth}|p{.2\textwidth}|}
		\hline
		Pixel array & 256 $\rm \times$ 256  \\
		\hline
		Pixel size & 55 $\rm \mu$m $\rm \times$ 55$\rm \mu m$ \\  
		\hline  
		Min. threshold &  $\rm \sim$ 500 $\rm e^-$ \\
		\hline
		Peaking time &  $<$ 25 ns \\  
		\hline  
		Time walk &  $<$ 25 ns \\
		\hline 
		Measurements & ToA \& ToT \\  
		\hline
		Count rate &  500 Mhit/s/chip \\
		\hline
		Readout &  Continuous, sparse, on-chip clustering \\  
		\hline  
		Output bandwidth & $>$ 12 Gbit/s \\
		\hline
		Power consumption & $<$ 3 W/chip \\  
		\hline  
		Radiation hardness & $>$ 500 MRad, SEU tolerant \\ 
		\hline    
      
      \end{tabular}      
      \end{center}
      \caption{Main VeloPix features}
      \label{tab:velopix}
      \end{table}

    \paragraph{Strip modules}
      The future VELO could keep the same tracking concept than the current, introducing improvements according to the evolution of the silicon technologies.
      The higher luminosity of the upgrade leads to increase occupancy.
      To deal with the it the granularity will be improved by increasing the number of strips, reducing the minimum pitch down to $\rm 30~\mu$m, and varying the pitch through the sensor to keep constant the occupancy per strip. 
      The planned sensors will be n-on-p type and $\rm 200~\mu$m thick.
      A new ASIC is under development exploiting synergies with other silicon detectors in LHCb, which will provide functionality on chip like clustering, common mode suppression and pedestal subtraction.
      The average rate per ASIC will be around 1.4~Gbits/s, and the data rate from the whole VELO will be above 2 Tbit/s.

      R and $\phi$ prototypes were already produced (Fig.~\ref{subfig:strip_prototype}) and they will be tested in the coming testbeams.
      
    \subsection{Mechanics and cooling}
    \begin{figure}[t]
    \centering 
    \subfigure[Metallized diamond]{
      \includegraphics[bb=0 0 125 110]{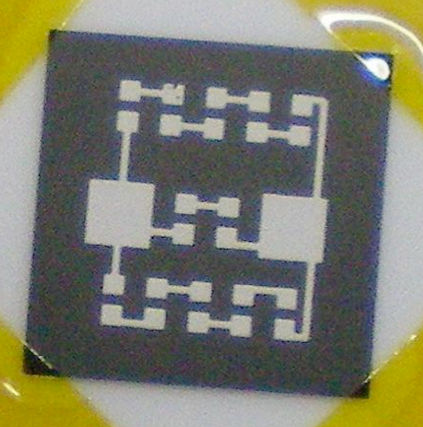}
      \label{subfig:diamond}
    }
    \subfigure[Microchannels etched onto a silicon substrate]{
      \includegraphics[bb=0 0 150 130]{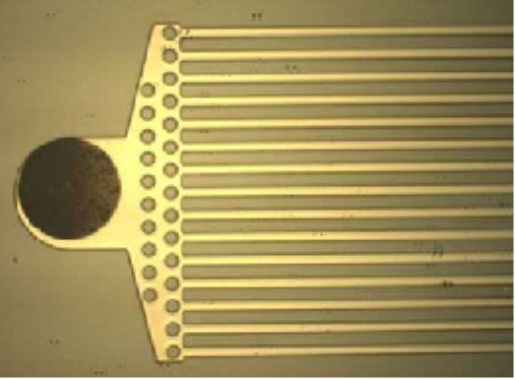}
      \label{subfig:microchannel}
    }
    \caption{\small Cooling options for the future VELO module}
    \end{figure}  
      The RF-foil is the aluminium box which encloses the VELO in a secondary vacuum.
      It allows overlap between the halves of the detector and shields the front end electronics from the RF noise pick-up and from beam currents.
      Depending on the module technology, a custom RF-foil must be manufactured. 
      A new technique will be used, milling the $\rm 250~\mu$m thick box from a solid aluminium block by a 5-axis milling head.
      The milling process is better than the pressing method used in the current VELO because it can manufacture sharp corners needed for the pixel option and allows a better thickness control.
      
      A common feature in strip and pixel-based modules will be the cooling spine.
      It will provide mechanical support but also cooling up to the very end of the sensor in order to avoid thermal runaway due to the increased occupancy.
      To cool down the cooling spine two main options are being developed, metallised CVD diamond and microchannel.
      In the first case, thermal conductivity and mechanical needs are guaranteed by a diamond spine clamped to a cooling block, while the IO signals of the ASIC will be carried out by a thermally activated silver paste deposited on the diamond (Fig.~\ref{subfig:diamond}).
      In the microchannel solution $\rm \sim200\times70~\mu m^2$ channels are etched onto a silicon substrate, where the $\rm CO_2$ is forced to pass through \citep{microchannels, 1748-0221-7-01-C01111}. 
      The layout of the channels can be adapted according to cooling needs. 
      Several prototypes were already produced and tested in laboratory environment (Fig.~\ref{subfig:microchannel}).
      
  \section{Tesbeam program}
  \begin{figure}[!t]
  \centering
  \includegraphics[bb=0 0 250 200]{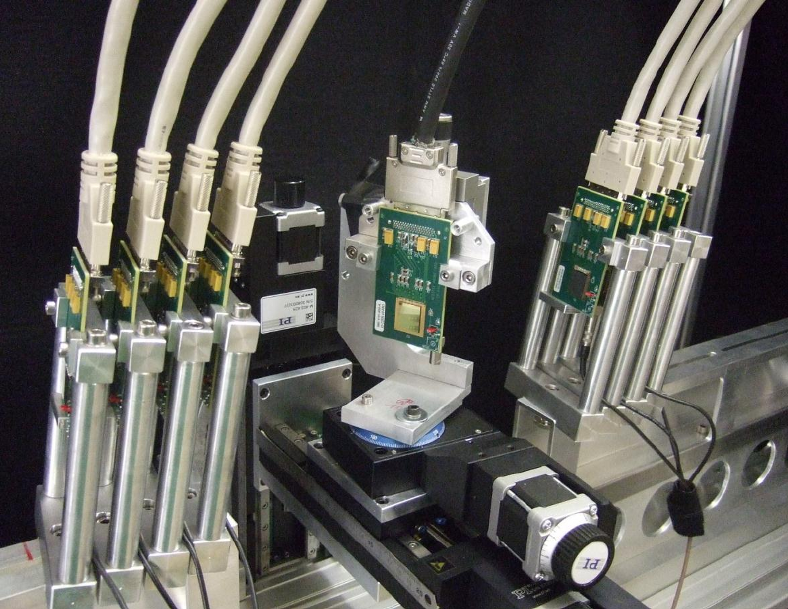}
  \caption{The Timepix based telescope}
  \label{fig:telescope}
  \end{figure}
  A key element of the testbeam program is the Timepix based telescope (Fig.~\ref{fig:telescope}).
  The telescope is divided in two arms, with four Timepix devices in ToT\footnote{\textbf{T}ime \textbf{o}ver \textbf{T}hreshold provides a value proportional to the deposited energy} mode allowing a track reconstruction with a spatial resolution $<~2~\mu$m at the Device Under Test (DUT) plane, with a $\rm \pi$ beam of 180 GeV/c.
  An additional Timepix device is placed at the end of the telescope and configured in ToA\footnote{\textbf{T}ime \textbf{o}f \textbf{A}rrival provides the time-stamping of the track} mode.
  The track rate goes up to 12 kHz depending of the DUT capabilities.
  The DUT is placed in between the two arms of the telescope, and it can be moved, rotated and cooled by a portable $\rm CO_2$ plant.
  
  During the 2012 testbeam campaign more than 30 different devices were tested to compare different options like performance after irradiation (from 0.5 to $\rm 2.5\times10^{15}\ 1MeV n_{eq}/cm^{2}$), silicon doping (n-on-p or n-on-n), edge distances, guard rings designs (from 0 to 6) and different vendors.  
  
  Detailed information of the testbeam program is available in \cite{dHynds}.
  
  \begin{figure}[hb]
  \centering
  \includegraphics[bb=0 0 220 175]{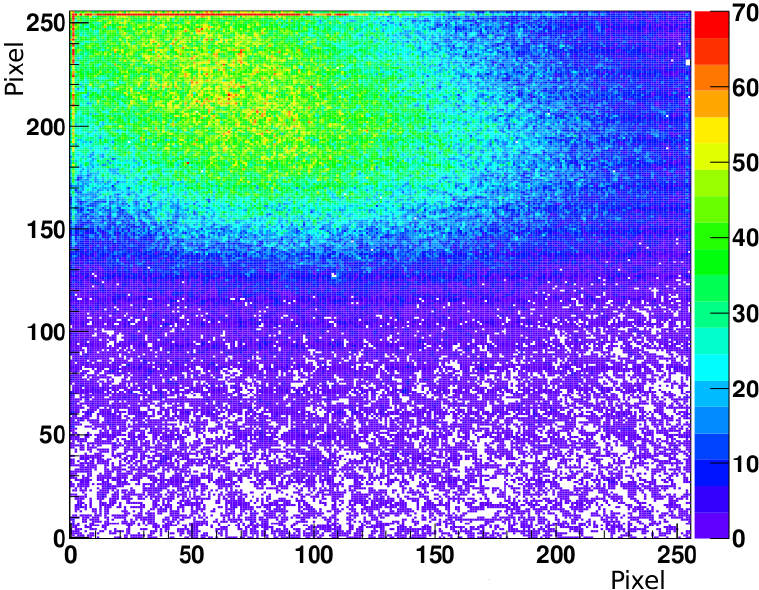}
  \caption{Beam hitmap in an edgeless Timepix device}
  \label{fig:results}
  \end{figure}
  \subsection{Results}
	Fig.~\ref{fig:results} shows a resulting plot from the 2012 testbeam campaign. 
	The device under test was an n-on-n edgeless sensor, 150~$\rm \mu$m thick, bump bonded to a Timepix chip and the beam incidence was perpendicular to the sensor.
	In the up left corner can be seen the beam spot. 
	In that region the pixels located in the second to last row have a higher occupancy. 
	This is a consequence of an edge effect, as in the last row the electric field is non uniform and drive the charges to the second row of pixels.

  \section{Schedule}
  The upgraded VELO will be installed during the Long Shutdown 2, starting in 2018. 
  A Letter of Intent \cite{LOI}, and a Framework Technical Design Report \cite{FTDR} were already published describing the LHCb Upgrade. 
  To be ready for the production stage, an intense R\&D program is underway providing useful information to choose the most suitable technology on the different aspects detailed previously.
  That decision will be made in the second quarter of 2013, when the Technical Design Report will be published.
  
  \section{Conclusions}
  The requirements for the LHCb VELO upgrade are very demanding.
  The radiation dose will be increased by a factor $\geq~10$ and the readout bandwidth will be increased by a factor of 40.
  R\&D effort is underway working in parallel solutions: pixel and strip based detector options are being developed, cooling solutions like metallised CVD diamond or microchannel are also being investigated. 
  To reduce the material budget in elements placed in the acceptance (modules, RF-foil) is another of the main concerns of the upgrade program.
  An intense testbeam program is being carried out to study sensor technologies, radiation hardness, cooling schemes and readout electronics.
  
  The list of requirements is not fully closed as we wish to get even closer to the beam, although it will increase considerably the readout rate.
  
  And everything with the aim of achieving, if not improving, the excellent resolution and efficiency of the current VELO.





\bibliographystyle{model1-num-names}
\bibliography{RESMDD12_VELO_Upgrade}







\end{document}